\documentclass[aps,prl,floatfix,epsfig,twocolumn,showpacs,preprintnumbers]{revtex4}
\usepackage{amsmath}
\usepackage{graphicx}
\usepackage{dcolumn}
\usepackage{bm}

\begin{document}

\title{Metallic ferroelectricity induced by anisotropic unscreened Coulomb
interaction in LiOsO$_{3}$}
\author{H. M. Liu$^{1}$, Y. P. Du$^{1}$, Y. L. Xie$^{1}$, J. -M. Liu$^{1}$,
Chun-Gang Duan$^{2}$ and Xiangang Wan$^{1}$}
\thanks{Corresponding author: xgwan@nju.edu.cn}
\affiliation{$^{1}$National Laboratory of Solid State Microstructures and Department of
Physics, Nanjing University, Nanjing 210093, China\\
$^{2}$Key Laboratory of Polar Materials and Devices, Ministry of Education,
East China Normal University, Shanghai 200062, China}

\begin{abstract}
As the first experimentally confirmed ferroelectric metal, LiOsO$_{3}$ has
received extensive research attention recently. Using density-functional
calculations, we perform a systematic study on the origin of the metallic
ferroelectricity in LiOsO$_{3}$. We\textbf{\ }confirm that the ferroelectric
transition in this compound is order-disorder like. In addition, we find
that the distribution of the conduction electron is extremely anisotropic,
which suggests that the electric screening ability is also highly
anisotropic. Then, by doing electron screening analysis, we unambiguously
demonstrate that the long range\emph{\ }ferroelectric order in LiOsO$_{3}$
results from the incomplete screening of the dipole-dipole interaction along
the nearest neighboring Li-Li chain direction. We therefore conclude highly
anisotropic screening and local dipole-dipole interactions are two most
important keys to form LiOsO$_{3}$-type metallic ferroelectricity.
\end{abstract}

\date{\today }
\pacs{71.20.-b, 72.80.Ga, 77.80.B-, 61.50.Ah}
\maketitle

The ferroelectric (FE) instability can be explained by a delicate balance
between short-range elastic restoring forces supporting the undistorted
paraelectric (PE) structure and long-range Coulomb interactions favoring the
FE phase \cite{fe1}. Itinerant electrons can screen the electric fields and
inhibit the electrostatic forces, metallic systems are thus not expected to
exhibit ferroelectric like structural distortion. Despite the
incompatibility, using a phenomenological theory, Anderson and Blount
proposed in 1965\emph{\ }that metals can break inversion symmetry \cite%
{Anderson}. They found that the FE metal is possible through a continuous
structural transition accompanied by the appearance of a polar axis, and the
disappearance of an inversion center \cite{Anderson}. In 2004, Cd$_{2}$Re$%
_{2}$O$_{7}$ had been proposed as a rare example of ferroelectric metals
\cite{cro1}, however, it was found that although this compound exhibits a
second order phase transition to a structure that lacks inversion symmetry,
a unique polar axis could not be identified \cite{NM-news}, which does not
fit the criteria about the FE metal.

Very recently, the first convincing success was achieved experimentally in
LiOsO$_{3}$ \cite{lso}. LiOsO$_{3}$ remains metallic behavior while it
undergoes a second-order phase transition from the high temperature
centrosymmetric $R\overline{3}c$ to a FE-like $R3c$ structure at $T_{s}$%
~=~140K \cite{lso}. Neutron and x-ray diffraction studies showed that the
structural phase transition involves the displacements of Li ions
accompanying also a slight shift of O ions \cite{lso}. The electronic
structure and lattice instability were studied by several groups \cite%
{lso1,lso2,lso3}. It was found that the local polar distortion in LiOsO$_{3}$
is solely due to the instability of the \textit{A}-site Li ion \cite%
{lso1,lso2,lso3}. The importance of the Coulomb interaction among 5$d$
electrons and the hybridization between oxygen $p$ orbitals and Os empty $%
e_{g}$ orbitals has also been emphasized by Giovannetti and Capone\cite{lso3}%
. Despite these efforts devoted to understanding the origin of the FE like
structural transition in this metallic system, there are still two
fundamental issues have not been clearly clarified. The first is the origin
of the ferroelectric instability: is it displacive or order-disorder?
Second, as FE-like phase transition of LiOsO$_{3}$ occurs at a relatively
high temperature (140 K), how can these local dipoles lined up to form
long-range order, as if there is no conduction electrons to screen the
dipole interactions.

In this paper, based on the density functional theory (DFT) calculations, we
reveal the microscopic mechanism for the FE-like structural transition in
LiOsO$_{3}$. Our study shows that different from other 5$d$ transition metal
oxides \cite{soc1,soc2,soc3,ee,nso1,nso2,nso3}, the spin-orbital coupling
(SOC) and the electronic correlation do not play important role in LiOsO$%
_{3} $. Our comprehensive potential surface calculations suggest that the
structural transition is order-disorder like. The most striking finding is
that the electric screening in LiOsO$_{3}$ is highly anisotropic despite its
metallic nature. Consequently, the dipole-dipole interactions are unscreened
along certain directions, which results in the long range FE order at
considerably high temperature. This is in sharp contrast to the case in the
displacive type FE compounds, where the FE structural transition is usually
driven by hybridization or lone-pair \cite{khomo} and consequently the
change of electric dipole (namely the atomic motion) will modify the valence
band significantly. If such displacive type FE compounds become metallic,
the interactions between their electric dipoles will be strongly screened
out, and the metallic FE phase is highly unlikely to occur.

Before the formal presentation of the calculated results, we would like to
first discuss our strategy to study the electric screening effect. As is
well known, the major difference between the insulator and metal is that
there are free non-localized electrons in metals, whereas in insulators
there are only bound electrons. Consequently, electrostatic forces will be
strongly screened by the itinerant electron in the metallic system. The
screening effect actually can be described as the electron charge difference
induced by a perturbation such as the change of the dipole or external
electric field \cite{DuanPRL}. Yet seldom efforts have been carried out to
study the screening effect in the bulk metal, as people generally believe
there is no macroscopic electric field inside metals. But in current study
we will control the formation of local dipoles in the bulk metal and study
the charge difference (response) caused by the change of dipole. This
provides a explicit picture on the exact behavior of the screening effect in
metallic systems.

\begin{figure}[tbph]
\centering
\includegraphics[width=8 cm]{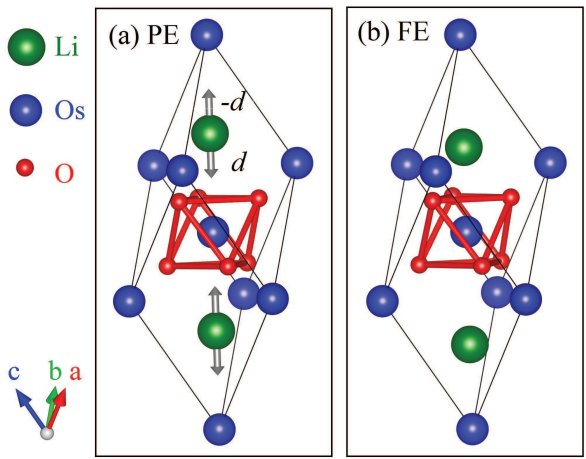}
\caption{(Color online) Primitive unit cell of (a) PE and (b) FE phases of
LiOsO$_{3}$. The green, blue and red balls are the Li, Os and O ions
respectively. $d$ and $-d$ corresponds to the displacements of Li ions along
the polar axis. }
\label{fig:struc}
\end{figure}

Our first-principles calculation is performed using the VASP (Vienna
ab-initio simulation package) code \cite{vasp1,vasp2}. The results presented
in the following are obtained by using the GGA (generalized gradient
approximation) PBE (Perdew-Becke-Erzenhof) function \cite{gga2}, a 20$\times
$20$\times $20 mesh for the Brillouin-zone sampling and 500 eV for cut-off
of the plane-wave basis set. The effect of exchange-correlation function,
pseudopotential, cut-off value % for basis set
has been carefully checked, and some of the results are shown in the
Supplementary Materials \cite{supple}.
\begin{table}[tbp]
\begin{ruledtabular}
\begin{tabular}{ccccc}
      & Atom & Position \\   \hline
    & Li & (0.25, 0.25, 0.25)  \\
  PE & Os & (0,0,0) \\
    & O & (0.8798, -0.3798, 0.25)  \\  \hline
    & Li & (0.2147, 0.2147, 0.2147) \\
 FE(Expt.) & Os & (0,0,0) \\
    & O & (0.8785, -0.3837, 0.2627)  \\ \hline
        & Li & (0.2149, 0.2149, 0.2149) \\
 FE(Calc.) & Os & (0,0,0) \\
    & O & (0.8855, -0.3842, 0.2557) \\
  \hline
\end{tabular}
\end{ruledtabular}
\caption{Atomic positions (in primitive rhombohedral coordinates) in PE and
FE phases. The experimental results are from Ref.~\protect\cite{lso}.}
\label{tab1}
\end{table}

There are 10 atoms in the primitive unit cell of LiOsO$_{3}$. The atomic
arrangements are sketched in Fig.~\ref{fig:struc}. In the $R\overline{3}c$
PE structure, the Os atoms are at the centers of the oxygen octahedrons,
while Li atoms are centered between two adjacent Os atoms along the polar
axis on average. Using the experimental lattice parameters, we optimize all
independent internal atomic coordinates of the FE structure until the
Hellman-Feynman forces on every atom are converged to less than 1 meV/{\r{A}}%
, the optimized internal atomic coordinates are listed in Tab.~\ref{tab1},
the experimental PE and FE structure had also been presented at Tab.~\ref%
{tab1} for comparison. The calculated results coincide with previous
experimental and calculated results \cite{lso,lso1,lso2,lso3}, the FE
structural phase transition mainly involves the displacements of Li atoms:
Li atoms shift along the polar axis about $d\sim $0.47{\r{A}} from the mean
positions of the PE phase (see grey arrow $d$ in Fig.~\ref{fig:struc}(a))
and O atoms slightly displace about 0.056{\r{A}\ }\cite{lso,lso1,lso3}.

\begin{figure}[t]
\centering
\includegraphics[width=7 cm]{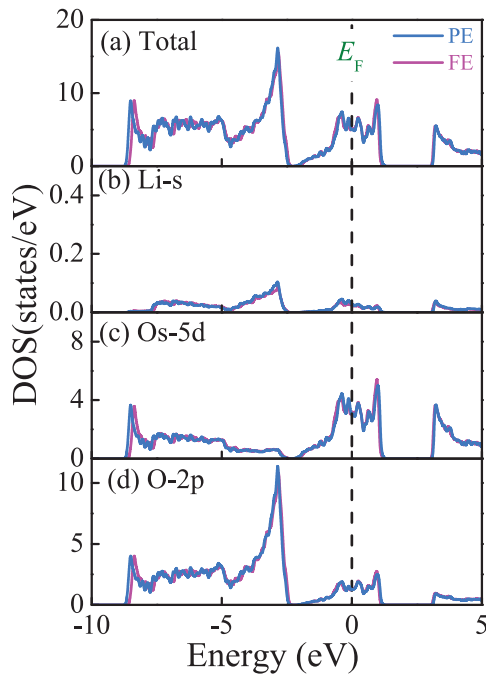}
\caption{(Color online) (a) The total DOS patterns of LiOsO$_{3}$ in
PE(blue) and FE(pink) phases. The partial DOS of (b) Li-1$s$ (c) Os-5$d$ and
(d) O-2$p$ states in PE(blue) and FE(pink) phases respectively. The Fermi
energy is positioned as zero. }
\label{figdos}
\end{figure}

Based on calculated lattice structure, we first perform standard GGA
calculation to see the basic features of the electronic structure of LiOsO$%
_{3}$. We show the total and partial density of states (DOS) in Fig.~\ref%
{figdos}. Our results are consistent with previous work \cite{lso,lso1}. The
energy range, -9.0 to -2.4 eV is dominated by O-2$p$ orbital with an
additional contribution from the Os-5$d$ state indicating hybridization
between them. Li is highly ionic and its bands are far from the Fermi level.
The Os atom is octahedrally coordinated by six O atoms, making the Os 5$d$
band split into the $t_{2g}$ and $e_{g}$ states, and the $t_{2g}$ bands are
located from -2.2 to 1.2 eV, as shown in Fig.~\ref{figdos}(c). Due to the
extended nature of 5$d$ states, the crystal splitting between $t_{2g}$ and $%
e_{g}$ states is large, and the $e_{g}$ states are located about 3.0 eV
higher than the Fermi energy and disperse widely. As shown in the comparison
of DOS of PE and FE, the electronic structures almost do not change during
the phase transition, which is consistent with the previous theoretical work
\cite{lso1}. It is worthy to mention that this is quite different from
prototype FE systems such as BaTiO$_{3}$, in which hybridization is
necessary for the FE phase transitions \cite{bto1,bto2,bto3}. Thus we think
the hybridization is not the driving force for the structural instability in
LiOsO$_{3}$.

It is well known that the SOC of 5$d$ electrons is very strong \cite{soc}
and usually changes the 5$d$ band dispersion significantly, as demonstrated
in Sr$_{2}$IrO$_{4}$ \cite{soc1}, pyrochlore iridates and spinel osmium \cite%
{soc2,soc3}. In the case of LiOsO$_{3}$, as shown in Fig.~\ref{figdos}, the
O-2$p$ orbitals are almost fully occupied, while the bands of Li are mainly
empty, thus Os occurs in its 5+ valence state and there are basically three
electrons in its $t_{2g}$ band. Since $t_{2g}$ band is half filled, it is
natural to expect the effect of SOC to be small despite the large strength
of SOC \cite{nso1}. This has been confirmed by the comparison of the band
structures obtained in the presence and absence of SOC (see Fig.S~1 in
Supplemental Material \cite{supple}).

Although the 5$d$ orbitals are spatially extended, it has been found that
the electronic correlations are important for 5$d$ transition metal oxides
\cite{soc1,soc2,soc3,ee,nso1}. Here, we estimate the Sommerfeld coefficient
based on the numerical DOS at Fermi level. Our numerical result ($%
6.1~mJ~mol^{-1}~K^{-2}$) is just slightly less than that of experimental one
($\gamma =7.7~mJ~mol^{-1}~K^{-2}$) \cite{lso} which indicates the electronic
correlation is not important for LiOsO$_{3}$. This is in sharp contrast with
the iridates, spinel osmium \cite{soc1,soc2,soc3}, and NaOsO$_{3}$ \cite%
{nso1}.

\begin{figure}[t]
\centering
\includegraphics[width=8.5 cm]{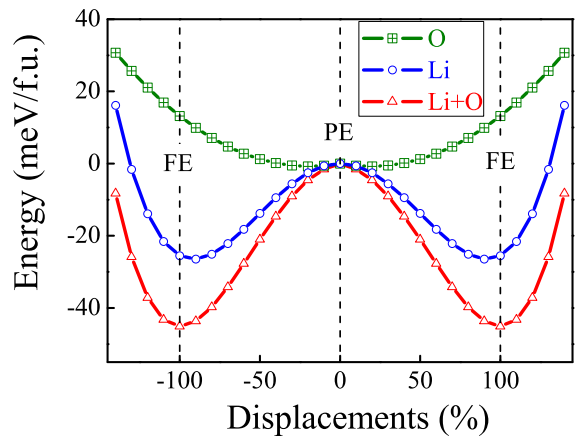}
\caption{(Color online) The olive, blue and red curves represents the
potential energy changes with respect to O displacements only, Li
displacements only and the coupled displacements of the Li and O ions. The
total energy and displacements of PE states are set as zero. The
displacements of corresponded FE states are set as 100\%.}
\label{figenergy}
\end{figure}

One fundamental issue about this system is the mechanism for the
ferroelectric instability, is it displacive or order-disorder? Using
comprehensive total energy calculations, we now try to solve this issue.
Following the common procedure used in the study of FE structures, we first
calculate the potential energy profile along different displacive soft
modes, i.e the evolution paths from the PE to FE structures. The results, as
shown in Fig.~\ref{figenergy}, suggest that the energy difference between
the PE and FE structures is majorally contributed by Li ions movements. The
depth of double wells resulting from the motion of Li ions only and both of
the Li and O ions are 27 and 44 meV, respectively, which is in consistent
with several previous works \cite{lso1,lso3}. Note that we find it is
important to adopt the optimized structure to obtain the correct potential
energy surfaces. Adopting experimental coordinates will obtain unreasonable
results, as the well depth caused by the sole Li ion movements is even
larger than that of considering both the Li and O moments (see Fig.S2 in
Supplemental Material \cite{supple}).

Nevertheless, we notice that the experimental transition temperatures of
LiOsO$_{3}$ is\textbf{\ }140 K and much lower than the depths of double
wells. This indicates the transition in LiOsO$_{3}$ is most probably
order-disorder like \cite{lno1}. This may also explain the experimentally
observed incoherent charge transport above the transition temperature \cite%
{lso}, which is possibly caused by the scattering induced by disorder of Li
off-center displacement. For order-disorder transition, the Li atoms
oscillate between the double wells and the potential\ wells remain basically
unchanged throughout the phase transition, thus we expect there is no
softening mode in the Raman spectra of LiOsO$_{3}$.

\begin{figure}[t]
\centering
\includegraphics[width=8 cm]{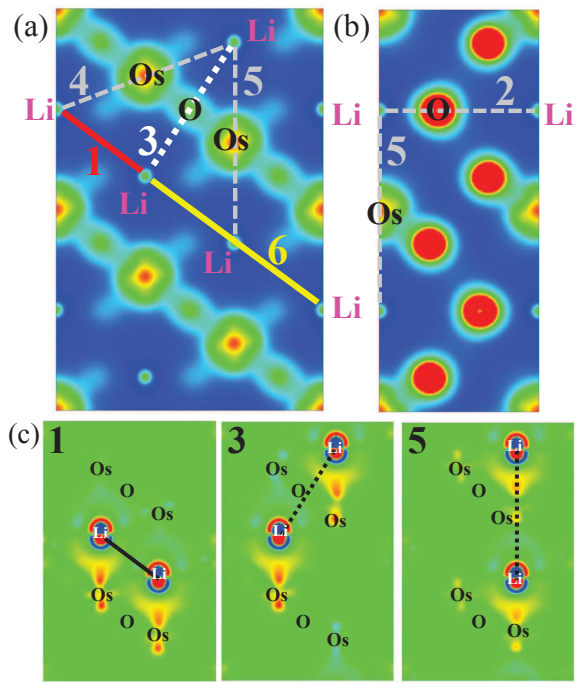}
\caption{(Color online) Partial electron densities contour maps for LiOsO$%
_{3}$ taken through (a) [1 -1 0] and (b) [2 -1 0] plane. Contour levels
shown are between 0 (blue) and 0.3 e/{\AA }$^{3}$ (red). (c) Charge density
difference between FE and PE structures for Li pair 1, 3 and 5 through [1 -1
0] plane. See text for details. Contour levels shown are between -0.004
(blue) and 0.004 e/{\AA }$^{3}$ (red).}
\label{figcharge}
\end{figure}

As mentioned above, the Li ions in LiOsO$_{3}$ favor an off-center
displacement and form local dipoles as shown in Fig.~\ref{fig:struc}. Thus
it is a puzzle why the local electric dipoles in different unit cell can
interact with each other and form a long range order at 140 K, noticing that
the distance between them is far (even the nearest neighbor dipole distance
is larger than 3.5 {\AA }) and the DOS at the Fermi level is rather large
(Fig.~\ref{figdos}). We find that the bands located below -10 eV are quite
narrow and have negligible hybridization with other bands. The electrons at
these bands are tightly bounded with the ion, thus almost do not change with
the motion of Li ion, namely these electrons almost have no contribution to
the electric screening effect. On the other hand, the displacements of Li
ions just slightly affect the Os-5$d$ and O-2$p$ electrons as shown in Fig.~%
\ref{figdos}. To have a straightforward view of the charge distribution, we
sketched the electron densities arising from states between -10 eV and the
Fermi level in Fig.~\ref{figcharge}(a,b). There are two distinctive
characters in these two figures. One is that the electronic density is
relevant high between the Os and O ions, which again indicates the strong
hybridization between Os-5$p$ and O-2$p$ states. The second is that there is
almost no conduction charge at all in a relative large space around the Li
ions, i.e. the Li ion is literally \emph{bare} ion. We will demonstrate
later that the later character directly results in incomplete electric
screening of dipole-dipole interactions and forms long range dipole ordering.

Following the previously described procedure, we then demonstrate the
screening effect by doing charge difference calculation. Since the FE like
transition basically involves displacements of Li ion, the change of local
dipole can be approximated by the Li movement from the PE structure and the
dipole interactions can be labeled as Li-Li pairs. Fig.~\ref{figcharge}, we
use symbol 1, 2, 3 etc. to denote the Li-Li pairs with the first, second and
third nearest distance between them. As is clear in Fig.~\ref{figcharge}(a),
there is almost no conduction charge distribution between pair 1 (red solid
line), thus it is natural to expect the screening effect for the nearest
dipole-dipole interaction is small. This had been confirmed by the
calculation of the charge difference induced by the Li motions. We first
construct a 3$\times $3$\times $1 supercell containing 270 atoms in
hexagonal phase. The PE structure is taken as background. Then we move Li
ions in pair 1 to their FE positions, and all other atoms are fixed. In this
way we can see what happened when the two local dipoles are formed. The
charge density change (namely $\Delta \rho =\rho _{FE}-\rho _{PE}$) is then
plotted in the left panel of Fig.~\ref{figcharge}(c), which clearly
demonstrates the screening of these two dipoles. As can be seen from this
plot, significant conduction electron responses only occur around Os ions,
and they form local dipoles to against the Li dipoles, this is exactly the
screening effect which is expected in metallic systems. Whereas the charge
distribution near O remains almost unchanged, indicating the Os-5 $d$ and O-2%
$p$ hybridization is neither important for the electric dipole interaction,
nor will be affected by the Li dipoles. The most interesting thing is, as
shown in the left panel of Fig.~\ref{figcharge}(c), there is almost no
modification of charge distribution between pair 1 at all. This clearly
demonstrates that dipole interaction between pair 1 is only slightly
screened. We then apply the same strategy to study other Li pairs. For pair
2 (Fig.~\ref{figcharge}(b)), as there is O atom between two Li ions, we
observe noticeable screening to prevent the direct dipole interaction. Yet
for pair 3, the dipole interaction is again not fully screened as shown in
the middle panel of Fig.~\ref{figcharge}(c). Detailed analysis indicates
that this is because this Li pair is 0.65 {\AA } away from the O ion plane.
For other pairs, as there is either O or Os atoms between two Li ions, the
electric screening effect is strong. The example of pair 5, which is also at
the [1 -1 0] plane like pair 1 and 3, is shown in the right panel of Fig.~%
\ref{figcharge}(c) for comparison.

\begin{table}[tbp]
\begin{ruledtabular}
\begin{tabular}{ccccccc}
   NN & 1st & 2nd & 3rd & 4th & 5th & 6th \\   \hline
  $d_{i}$ ({\AA}) & 3.66  & 5.06 & 5.28 & 6.25 & 6.61 & 7.32 \\
  $J_{i}$ (meV) & -4.2 & -0.16 & -1.9 & -0.17 & -0.06 & -0.27 \\
\end{tabular}
\end{ruledtabular}
\caption{The distances $d_{i}$ ($i$=1-6) and coupling parameters $J_{i}$ ($i$%
=1-6) between $i$th Li ion pairs.}
\label{tab2}
\end{table}

Above we have provided a qualitative picture about why locale dipole
interactions are not fully screened in LiOsO$_{3}$. To give a more
quantitative explanation, we try to estimate the interaction strength
between the local electric dipole moments. Again using the above adopted
supercell, we obtain coupling constant $J_{i}$ ($i$=1-6) between the $i$ th
Li pairs from the energy difference between the local FE\textbf{\ }and
antiferroelectric (AFE) states ($i$ th Li pairs are AFE ordered), i.e $J_{i}=
$[E$_{FE}$-E$_{AFE}$]/2. The Li-Li distances $d_{i}$ of each pair and the
obtained interaction parameter $J_{i}$ are listed in Table II.
Consistent with the above screening discussions, $J_{1}$ and $J_{3}$ are
much larger than all of other interactions, indicating the dipole
interactions are highly anisotropic. Despite $d_{6}$ is longer than $d_{2}$,
$d_{4}$ and $d_{5}$, $J_{6}$ is considerably larger than $J_{2}$, $J_{4}$
and\ $J_{5}$ as shown in Table II, which also indicate the anisotropic
screening effect in this metallic compound.

To show these dipole interaction parameters obtained in above procedure are
reasonable, we perform the Monte-Carlo (MC) simulations using an effective
Ising-like Hamiltonian: $H=\sum_{i}J_{i}D_{m}D_{n}$, $J_{i}$ is the coupling
constant between dipole moments $D_{m}$ and $D_{n}$. The obtained phase
transition temperature is 210 K with only $J_{1}$ considered and 330 K with
both $J_{1}$ and $J_{3}$ considered, which is reasonably higher than the
experimental $T_{s}$ (140 K), and the overestimation may come from the rigid
dipole model used in our MC simulations. This, again, show that our
explanation on the mechanism of the lineup of local dipoles in metallic LiOsO%
$_{3}$ is self consistent.

In summary, we use a straightforward way to study the electric screening
effect in bulk metallic systems. We conclude that microscopic mechanism of
metallic ferroelectricity in LiOsO$_{3}$ is not related spin-orbital
interaction or electronic correlation, as one may expect for 5\textit{d}
system. It is the highly inhomogeneous charge distribution in this metallic
system plays crucial roles in the ferroelectric-like structural transition.
The highly ionized Li atoms are easy to form local dipoles as a consequence
of the Coulomb interactions. The screening in this compound, different from
our previous thought for a metallic system, is strongly anisotropic.
Therefore local dipole interactions along certain directions are not fully
screened and result in the dipole ordering below rather high temperature. We
also want to emphasis that the above picture implied that the structural
phase transition should be order-disorder instead of displacive type, which
is also supported by our potential energy surface calculations. This is
because the displacive-type ferroelectric transition occurring in ABO$_{3}$
perovskite structures are general triggered by the hybridization between the
B and O electronic states or the lone-pair in A site. In either case the
change of the electric dipole will modify the valence band considerably.
Therefore in displacive system the dipole-dipole interaction will be
strongly screened out in the metallic phase.

HML is thankful to Z. Z. Du and Y. L. Wang for helpful discussions. The work
was supported by the National Key Project for Basic Research of China
(Grants No. 2011CB922101, 2014CB921104), NSFC under Grants No. 91122035,
11174124, 51431006, 11374147, 51332006 and 61125403. The project is also
funded by Priority Academic Program Development of Jiangsu Higher Education
Institutions.

\section{Supplementary}
\begin{figure*}[t]
\centering
\includegraphics[width=15 cm]{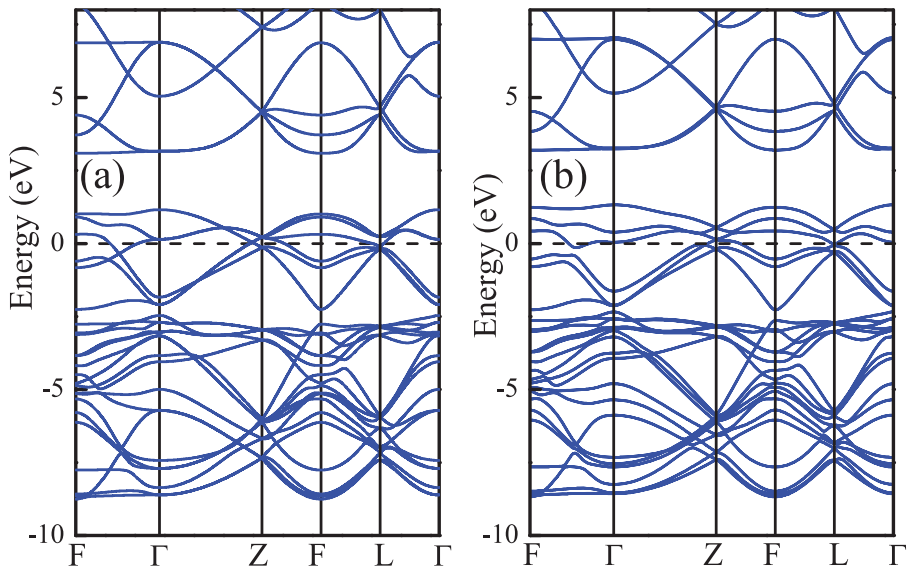}
\caption{Band structure of LiOsO$_{3}$, shown along the high-symmetry
directions. (a) GGA, (b) GGA + SO. }
\label{band}
\end{figure*}

In order to investigate the effect of spin-orbital coupling (SOC) on the
electronic structures, we compare the results obtained with and without SOC.
As demonstrated in Fig.~\ref{band}, the band structure difference around the
Fermi level is negligible.

%%%%%%%%%%%%%%%%%%%%%%%%%%%%%%%%%%%%%%%%%%%%%%%%%%%%%%%%%%%%%%%%%%%
%%%%%%%%%%%%%%%%%%%%%%%%%%%%%%%%%%%%%%%%%%%%%%%%%%%%%%%%%%%%%%%%%%%

We also calculate potential energy surface based on experimental
displacements \cite{lso} by using available exchange-correlation functionals
(such as: LDA, PBE and PW91) and the results are presented in Fig.~\ref%
{doublewell}. In each case, dense $k$ mesh and huge energy cutoff for the
basis set are carefully checked for better convergence. Adopting
experimental coordinates will obtain unreasonable results, as the well depth
caused by the sole Li ion movements is even larger than that of considering
both the Li and O moments.

%(experimental FE phases, 18 meV an illogical results shows  is shallower than that
%%%%%%%%%%%%%%%%%%%%%%%%%%%%%%%%%%%%%%%%%%%%%%%%%%%%%%%%%%%%%%%%%%%
\begin{figure}[t]
\centering
\includegraphics[width=16 cm]{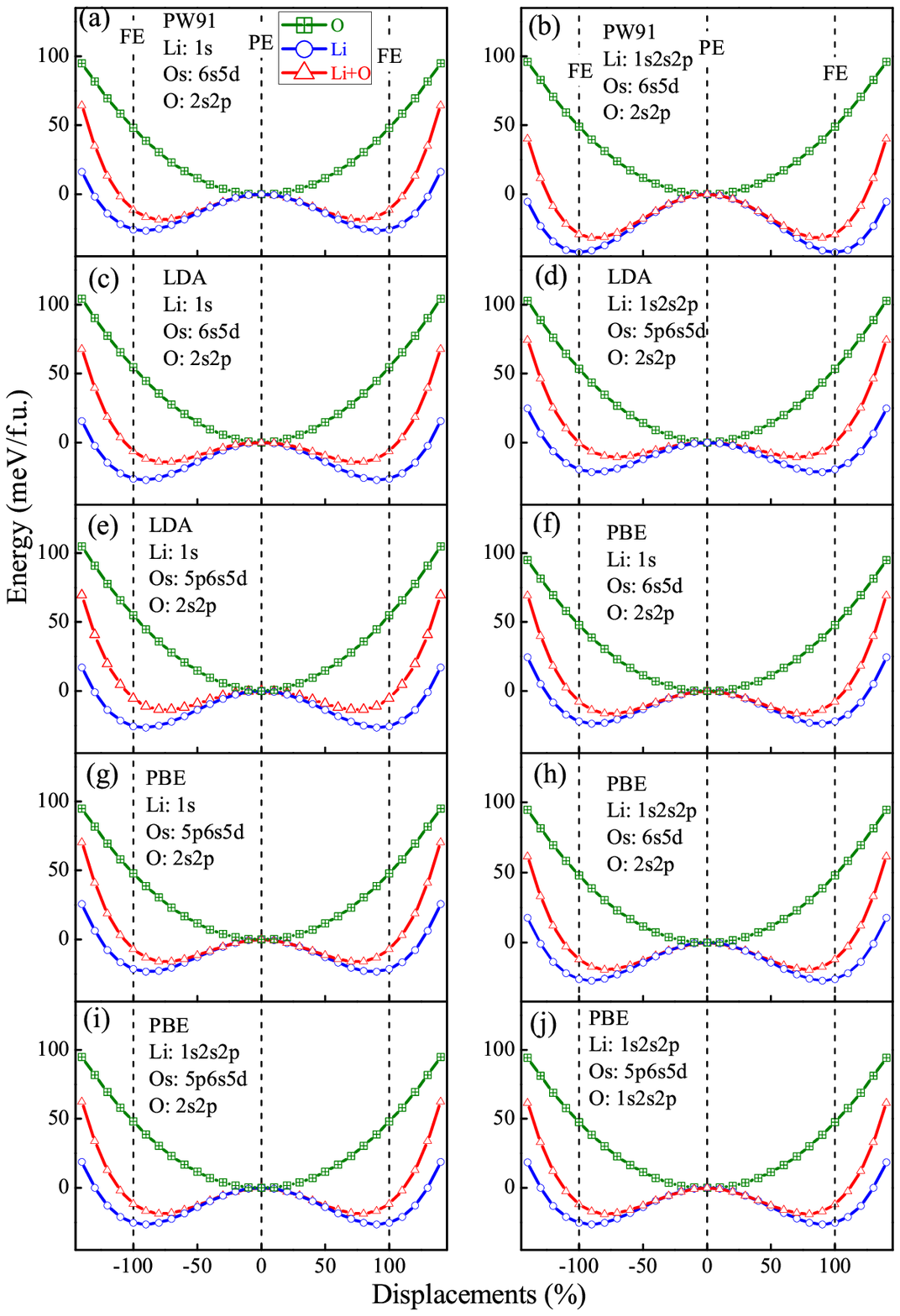}
\caption{(Color online)Potential energy surface based on experimental
displacements \protect\cite{lso} using other available exchange-correlation
functionals: (a)-(b) PW91, (c)-(e) LDA, (f)-(j) PBE. For each
exchange-correlation functional, the calculations are performed using
pseudopotentials with different valence electrons, and the used valence
states for Li, Os and O are inserted in related figures. The olive, blue and
red curves represent the potential energy changes with respect to O
displacements only, Li displacements only and the coupled displacements of
the Li and O ions. The total energy and displacements of PE states are set
as zero. The displacements of corresponded FE states are set as 100\%. }
\label{doublewell}
\end{figure}
%%%%%%%%%%%%%%%%%%%%%%%%%%%%%%%%%%%%%%%%%%%%%%%%%%%%%%%%%%%%%%%%%%%

\end{document}